\documentstyle[version2,preprint,aps]{revtex}
\begin{document}
\preprint{UIUC-P-95-07-050}
\draft
\begin{title}
Transport through Dirty Luttinger Liquids Connected to
Reservoirs
\end{title}
\author{Dmitrii L. Maslov}
\begin{instit}
Department of Physics and Materials Research
Laboratory\\
University of Illinois at Urbana-Champaign,
Urbana, IL 61801, USA;\\and Institute for Microelectronics Technology,
Academy of Sciences of Russia\\ Chernogolovka, 142432 Russia
\end{instit}
\begin{abstract}
It is shown that the conductance of  a weakly disordered
Luttinger-liquid quantum wire connected to non-interacting
leads is affected  by electron-electron interactions in the wire.
This is in contrast to the case of a perfect wire the
conductance of which is given by $e^2/h$ regardless of interactions in the
wire. The disorder-induced correction to the conductance
scales with temperature and/or the wire length, the
scaling exponent being determined
only by the interaction strength in the wire.
These results  explain recent experiments on quasi-ballistic
GaAs quantum wires.
\end{abstract}
\pacs{PACS numbers: 72.10.Bg, 73.20.Dx}
\narrowtext
Transport through quantum wires is commonly believed
to be strongly affected by electron-electron interactions. In particular,
if only one channel of transverse quantization is open and the electrons
are in the Luttinger liquid state \cite{haldane}, the conductance of a
perfect wire is expected to be $Ke^2/h$
\cite{apel,fisher-prl,fisher-prb,fukuyama},
where the parameter $K$ characterizes the sign and the strength
 of the interactions:
$K<1$ for repulsion; $K>1$ for attraction;
$K=1$ in the absence of interactions.
Moreover, in the case of repulsion, disorder in the wire  is predicted
to suppress the conductance much more strongly than in the non-interacting case
\cite{apel,fisher-prl,fisher-prb,fukuyama,furusaki,matv_yue}.
This suppression comes about via temperature-  and/or
length-dependent corrections to the conductance
of a perfect wire that
diverge in the limit $T\to 0$ or $L\to \infty$
even for an arbitrarily weak disorder, indicating
the tendency to an insulating behavior.
The exponents
of the $T$- and $L$- scalings are also determined by the parameter $K$.
This has to be contrasted with the non-interacting theory, which
predicts $T$-independent conductance (at least
for $T$ smaller than the interband energy spacing). The
physical explanation
of the interaction-induced breakdown of conduction
\cite{fisher-prb} is that the $2k_F$ backscattering
due to impurities stimulates the divergence of charge-density-wave
fluctuations with the period $2\pi/2k_F$.
With all these
dramatic differences between the interacting and the non-interacting theories,
an experiment seems capable of discriminating between the two.

At the first sight, a recent experiment  on high-mobility
GaAs quantum wires \cite{tarucha}
has given support to neither theory, however.
Indeed, at higher temperatures ($T>1.2$K) the observed value of the
conductance was  $e^2/h$ with a high accuracy, which seems to support the
non-interacting theory. On the other hand, reduction
of the conductance was observed at lower $T$, and by fitting the observed
$T$-dependence of the conductance into the
interacting theory \cite{fukuyama} the authors of  Ref.~\cite{tarucha}
obtained a value of $K\approx 0.7$. This seems to support the
interacting theory, but also implies that at higher
$T$, where the disorder-induced correction is small,
the reduction of the conductance should have been of the order of $30\%$,
which clearly contradicts to the data.

A partial resolution to this paradox has been given in two recent papers
\cite{safi,maslov}, in which it was
emphasized  that the model of a homogeneous Luttinger liquid leading to
the result  $Ke^2/h$ of
Refs.~\cite{apel,fisher-prl,fisher-prb,fukuyama} is not adequate to a
typical  experimental situation. Indeed,
a narrow wire is always connected to wide electron reservoirs
by the conducting leads. As the leads and the reservoirs are necessarily
{\it not} one-dimensional, the electrons there form not a Luttinger
liquid but rather a Fermi liquid which, for the present purposes,
can be considered as a non-interacting Fermi-gas. Consequently, the effective
one-dimensional
model considered in Refs.~\cite{safi,maslov} was that of an inhomogeneous
Luttinger liquid  with the interaction strength varying from some non-zero
value in the
central part of the system (the \lq\lq wire \rq\rq\ ) to zero in the outer
parts (the \lq\lq leads\rq\rq\ ). It has been shown
that in the absence of disorder
the conductance of such a system is given by
$e^2/h$ {\it regardless} of the interactions in the wire.
This explains the absence of the conductance-renormalization at higher
$T$ observed in Ref.~\cite{tarucha}.
It must be added that remarks supporting this result
were made earlier by Kane and Fisher \cite{fisher-prl} and by Matveev
and Glazman \cite{matveev}.

The question that still remains open is whether the
experimentally observed $T$-dependence of the conductance
is a signature of a dirty Luttinger-liquid
state in the wire itself or is a reflection of
the presence of the leads as well.
In this paper, I show that the
$T$- and $L$-dependent corrections to the dc conductance of a weakly
disordered wire are determined entirely by the interactions in the wire,
and are not affected by the presence of the non-interacting leads. This
result, taken together with the results of Refs.~\cite{safi,maslov}, suggests
that the experimental observations of Ref.~\cite{tarucha} can be seriously
considered as an indication of the Luttinger-liquid state in GaAs
quantum wires.

The result announced above can be entirely anticipated from the following
simple
physical picture. Consider a weakly disordered wire containing a Luttinger
liquid and adiabatically connected to the Fermi-liquid leads.
 In the absence of disorder, the finite resistance
of a perfect wire ($=h/e^2$) is entirely due to contact resistance
\cite{imry,landauer,glazman,levinson}: some of the electrons coming from
the wide leads are reflected as the channel becomes narrower. This
reflection takes place outside the wire, where the electrons are in the
Fermi-liquid state, therefore, the contact resistance is not affected by
the interactions in the wire. Weak disorder in the wire
gives rise to an additional contribution to the
resistance. This contribution is determined by the scattering {\it in} the
wire, where electrons are in the Luttinger-liquid state, and therefore this
contribution has features typical of a Luttinger liquid but not of a
Fermi liquid.

As in Refs.~\cite{safi,maslov}, I consider an infinite Luttinger liquid
separated into three regions: the wire ($-L/2<x<L/2$) and the leads
($x\geq L/2$). The interaction parameter $K$ changes
abruptly from the value $K_{\rm W}$ in the wire to the value $K_{\rm L}$ in the
leads. Having in mind the experimental system studied in Ref.~\cite{tarucha},
I shall focus on the case of repulsive interaction in the wire
(i.e., $K_{\rm W}\leq 1$) and
put $K_{\rm L}=1$ at the end of the
calculation. The length $L$ corresponds to the length
of that segment of the original system where the
applied electrostatic potential varies most appreciably. Consequently, the
electric field is assumed to be zero for $|x|\geq L/2$. I also assume
that a random potential $V(x)$ is present only in the wire, i.e., for
$|x|<L/2$, and is weak
enough to be treated via perturbation theory \cite{fn_perturb}.
The current $I=ej$ is related to the electric field by
\begin{equation}
I(x,t)=\int^{L/2}_{-L/2}dx'\int \frac{d\omega}{2\pi}
 e^{-i\omega t}\sigma_{\omega}(x,x'){\bar E}_{\omega}(x'),
\label{eq:current}
\end{equation}
where ${\bar E}_\omega(x)$ is temporal Fourier component of the
electric field at point $x$
and $\sigma_{\omega}(x,x')$ is the non-local ac conductivity, which is given
by the Kubo formula \cite{shankar}
\begin{equation}
\sigma_{\omega}(x,x')=\frac{ie^2}{
\omega}\int^{\beta}_{0}\langle
T_{\tau}^{*} j(x,\tau)j(x',0)\rangle
 e^{-i{\bar \omega}\tau}\Big|_{{\bar\omega}=i\omega
-\epsilon}.
\label{eq:kubo}\end{equation}
(I put $\hbar=1$). In the bosonized
form, the particle-number current is given by
$j=-i\partial_{\tau}\phi/\sqrt{\pi}$
and Eq.~(\ref{eq:kubo}) reduces to \cite{shankar}
\begin{mathletters}
\begin{eqnarray}
\sigma_{\omega}(x,x')&=&e^2\frac{i{\bar\omega}^2}{\pi\omega}G_{{\bar
\omega}}(x,x')|_{{\bar\omega}\to i\omega-\epsilon},
\label{eq:kubo1}\\
G_{{\bar \omega}}(x,x')&=&\int^{\beta}_{0}d\tau\langle
T_{\tau}^{*} \phi(x,\tau)\phi(x',0)\rangle e^{-i{\bar \omega}\tau},
\label{eq:prop_def}\end{eqnarray}
\end{mathletters}where $G_{{\bar \omega}}$ is the
propagator of the boson field $\phi$.
The (Euclidean) action of the spinless Luttinger liquid is given
by $S=S_0+S_{\rm i}$,
where
\begin{equation}
\!\!\!\!S_{0}=\!\!\int\int^{\beta}_{0}dx\,d\tau\!\frac{1}{2K(x)}
\Big\{\frac{1}{v(x)}(\partial_{\tau}\phi)^2+v(x)(\partial_x\phi)^2\Big
\},
\label{eq:act_0}\end{equation}
$v(x)$ is the density-wave velocity and $S_{\rm i}$, which describes
backscattering due to disorder, is given by
\begin{equation}
\label{eq:act_i}
S_{\rm i}=\frac{2}{a}\int \! dx \!\int^{\beta}_{0}\!d\tau V(x)
\cos (2k_Fx+2\sqrt{\pi}\phi),
\end{equation}
in which $a$ is the microscopic length cut-off.
In the homogeneous model \cite{apel,fukuyama},
$\sigma_\omega$ in the presence disorder was found by using the Luther-Peshel
formula
\cite{luther}, which relates the effective mean free time to the $2k_F$
component of the density-density correlator. In the inhomogeneous
situation, the meaning of the mean free time is unclear, and I will
determine the
disorder-induced corrections to $\sigma_\omega$ directly
from the Kubo formula (\ref{eq:kubo}) via
perturbation theory in $S_{\rm i}$.
In what
follows, I will consider only the conductance averaged over the ensemble of
disorder realizations.
Assuming that $\overline{V(x)}=0$, where
$\overline{\mathstrut\dots}$ stands for the ensemble averaging,
the first non-vanishing correction to the ensemble-averaged propagator
is given by
\begin{eqnarray}
\overline{\delta G(X,X')}=\frac{1}{a^2}\!\!\int dX_1dX_2\overline{V(x_1)V(x_2)}
\cos(2k_{\rm F}(x_1-x_2))\nonumber\\
\!\!\!\!\left\{\langle \phi(X)\phi(X')Q(X_1,X_2)\rangle_0\!-\!
\langle \phi(X)\phi(X')\rangle_0\langle Q(X_1,X_2)\rangle_0\right\}
\label{eq:prop_corr}
\end{eqnarray}
where $X=\{x,\tau\}$,
$\langle\dots\rangle_0$ stands for averaging over Gaussian fluctuations
of $\phi$ with the weight $S_0$, and
\begin{equation}
Q (X_1,X_2)\equiv e^{i2\sqrt{\pi}\left[\phi(X_1)-
\phi(X_2)\right]}
\label{eq:gamma}
\end{equation}
At this stage, for the sake of simplicity I choose $V(x)$
in the form of white-noise: $\overline{V(x_1)V(x_2)}=
n_{\rm i}u^2\delta(x_1-x_2)$, where $n_{\rm i}$ is the concentration
of \lq\lq impurities\rq\rq\, and $u$ is the \lq\lq impurity
strength\rq\rq\rlap. The effective elastic mean free path $\ell$ (in the
absence of the
interactions) can then be defined as
 $1/\ell=n_{\rm i}u^2/a^2\omega_{\rm F}^2$,
where $\omega_{\rm F}$ is the (non-universal) ultraviolet energy cut-off
(of the order of the Fermi energy). Although in the real GaAs system
the impurity potential is long-ranged, the expectation is that this
simplification cannot significantly affect the final results,
provided that $\ell$ is replaced by the correct mean free path
for a more realistic disorder potential \cite{fn_mfp}.
Thus I find
the correction to the non-local conductivity:
\begin{eqnarray}
-\overline{\delta\sigma_{\omega}(x,x')}&=&\frac{2ie^2{\bar \omega}^2
\omega_{\rm F}^2}{\pi\ell\omega}
\int^{L/2}_{-L/2}d{\bar x}G_{{\bar \omega}}^0(x,{\bar x})G_{{\bar
\omega}}^0({\bar x},x')\nonumber\\
&\times &\left[F_{0}({\bar x})-F_{{\bar \omega}}({\bar x})\right]|_{{\bar
\omega}\to i\omega-\epsilon},
\label{eq:cond_corr}
\end{eqnarray}
where $G_{\bar\omega}^0$ is the propagator in the absence of
disorder and $F_{{\bar \omega}}(x)$ is the $\tau$-Fourier transform of the
(inhomogeneous) $2k_F$
density-density correlation function
\begin{eqnarray}
&F(x,\tau)\equiv \langle Q(x\tau,x0)\rangle_0\nonumber\\
&=\exp\Big[-\frac{4\pi}{\beta}
\sum_{{\bar \omega}}(1-e^{-i{\bar \omega}\tau})G_{{\bar \omega}}^0(x,x)\!\Big].
\label{eq:F}
\end{eqnarray}
The propagator
$G_{{\bar \omega}}^0(x,x')$ satisfies the equation
\begin{equation}
\bigg\{\frac{{\bar
\omega}^2}{v(x)K(x)}-\partial_x\Big(\frac{v(x)}{K(x)}\partial_x\Big)
\bigg\}G_{{\bar \omega}}^0(x,x')=\delta(x-x'),
\label{eq:prop}\end{equation}
and the following boundary conditions: (i) that $G_{{\bar \omega}}^0(x,x')$
be continuous
at $x=\pm L/2$ and $x=x'$; (ii) that
$\frac{v(x)}{K(x)}\partial_xG_{{\bar \omega}}^0(x,x')$ be
continuous at $x=\pm L/2$; but (iii) be discontinous
at $x=x'$ so that
\begin{equation}
-\frac{v(x)}{K(x)}\partial_xG_{{\bar \omega}}^0(x,x')\Big|^{x=x'+0}_{x=x'-
0}=1.
\label{eq:jump}\end{equation}
In addition, I assume that infinitesimal dissipation is present in
the leads, so that $G_{{\bar \omega}}^0(\pm\infty,x')=0$.
To evaluate the function $F$ in Eq.~(\ref{eq:F}), one needs to know
$G_{{\bar \omega}}^0(x,x')$ only for $-L/2\leq x=x'\leq L/2$. Straightforward,
albeit
lengthy,
algebra leads to the result
\begin{eqnarray}
\!\!\!\!G_{{\bar \omega}}^0=\frac{K_{\rm W}}{2|{\bar \omega}|}\!+\!\frac{K_{\rm
W}}{|{\bar \omega}|}\frac{\kappa_{-}^2e^{-L/L_{{\bar \omega}}}+
\kappa_{+}
\kappa_{-}\cosh(2x/L_{{\bar \omega}})}{e^{L/L_{{\bar \omega}}}
\kappa_{+}^2-e^{-L/L_{{\bar \omega}}}\kappa_{-}^2},
\label{eq:prop_xx}
\end{eqnarray}
where $L_{{\bar \omega}}\equiv v_{\rm W}/|{\bar \omega}|$, $v_{\rm W}$ is the
density-wave velocity in the
wire and $\kappa_{\pm}\equiv 1/K_{\rm W}\pm1/K_{\rm L}$. I now
consider separately the cases of \lq\lq high\rq\rq\
($v_{\rm W}/L\ll T\ll \omega_{\rm F}$)
and \lq\lq low\rq\rq\ ($T\ll v_{\rm W}/L$)
temperatures. (The quotations marks are intended to imply to that the
\lq\lq low\rq\rq\ temperature case can be alternatively viewed as the \lq\lq
long\rq\rq\ length
case and vice versa.)

At \lq\lq high\rq\rq\ temperatures, the second
term in Eq.~(\ref{eq:prop_xx}) is exponentially small ($\propto
\exp(-(L-2x)/L_{{\bar \omega}}$) unless ${\bar \omega}=0$. The term ${\bar
\omega}=0$ gives zero
contribution to the sum in Eq.~(\ref{eq:F}), however,
as can be seen by performing
the infrared regularization of the propagator
($|{\bar \omega}|\to\sqrt{|{\bar \omega}|^2+m^2}$) and then letting $m\to 0$.
Thus,
only the first term in Eq.~(\ref{eq:prop_xx}) has to be taken into account.
This term is precisely the same as in the case of a homogeneous
Luttinger liquid with parameter $K_{\rm W}$. Already at this stage it can be
anticipated
that the $T$-scaling of the conductance is determined by $K_{\rm W}$.
 The function $F$ is now $x$-independent and is given by
\begin{equation}
F=\Big[\frac{2\pi/\omega_F\beta}{\sin \pi\tau/\beta}\Big]^{2K_{\rm W}}.
\label{eq:F_high}
\end{equation}
After the analytic continuation  ${\bar \omega}\to i\omega-\epsilon$, the limit
$\omega\to 0$ can be taken.
In this limit  the remaining two propagators
in Eq.~(\ref{eq:cond_corr}) are $x$- and $x'$-independent: $\lim_{\omega\to
0}G_{\omega}^0(x,x')=iK_{\rm L}/2\omega$ \cite{maslov}. The rest of the
calculations
proceeds exactly as in the homogeneous case.
The resulting non-local dc conductivity is also $x$- and
$x'$-independent from which,
with the help of Eq.~(\ref{eq:current}), one finds that the dc conductance is
$g=\sigma_{0}$ \cite{schulz,maslov}. Recalling the result for the
conductance of a perfect wire, $g_{0}=K_{\rm L} e^2/h$ \cite{safi,maslov}, and
restoring $\hbar$,
the final result for the conductance can be written as
\begin{equation}
\!\!\!\overline{g}=g_0+\overline{\delta g}
=K_{\rm L}\frac{e^2}{h}\!-\!CK_{\rm L}^2\frac{L}{\ell}\Big[\frac{2\pi T}
{\hbar\omega_{\rm F}}\Big]^{2(1-K_{\rm W})},
\label{eq:res_high}
\end{equation}
where
\begin{equation}
C=8\sqrt{\pi}\sin(\pi K_{\rm W})\frac{\Gamma(1-K_{\rm W})}
{\Gamma(\frac{1}{2}+K_{\rm W})}\Big(\Gamma(K_{\rm W})\Big)^2.
\label{eq:const}
\end{equation}
Note that $K_{\rm L}$ enters only in the prefactors, whereas the exponent of
the $T$-scaling
is determined by $K_{\rm W}$. Tracing back through the calculations,
we can now see the reason
for this. The $K_{\rm L}$-dependence comes from the propagators
in the prefactor of Eq.~(\ref{eq:cond_corr}), which depend on the
frequency of the applied field $\omega$. In the limit $\omega\to 0$, they
become long-ranged and
contain only $K_{\rm L}$ but not $K_{\rm W}$. The
$T$-scaling comes the propagator entering the
function $F$ [Eq.~(\ref{eq:F})].
 This propagator does not depend on $\omega$. In the
\lq\lq\ high\rq\rq\ temperature limit, it becomes short-ranged, and
contains only $K_{\rm W}$ but not $K_{\rm L}$.

I now turn to the case of \lq\lq low\rq\rq\ temperatures. In previous
work dealing with the homogeneous model \cite{apel,fisher-prb,fukuyama}, this
case was treated by employing the Lorentzian invariance of the $(1+1)$D
field theory which ultimately reduces to a simple recipe: to obtain the result
at
\lq\lq low\rq\rq\ temperatures, one can  take the result for
\lq\lq high\rq\rq\ temperatures and replace $T$ by $v/L$, and vice versa.
This recipe works, however, only if $L$ has a
meaning of the total system size. In our inhomogeneous case, $L$ is the length
of the
wire which is only
the part of the system, the total system including both the wire and the
leads. Thus, the $T\Leftrightarrow v/L$ equivalence cannot {\it a priori} be
expected
to work. In fact, it will be shown below that such an equivalence exists
only if electrons in the leads do not interact, i.e., $K_{\rm L}=1$.

At \lq\lq low\rq\rq\ temperatures, the sum over
${\bar \omega}$ in Eq.~(\ref{eq:F}) can be replaced by the integral.
Performing the integration, I find
\begin{eqnarray}
&F(x,\tau)=\frac{1}{\Big[1+(\tau\omega_{\rm F})^2\Big]^{K_{\rm
W}}}\times\nonumber\\
&\!\!\!\exp\!\!\Big(\!\!-\!\!K_{\rm W} z_0\big[\Phi(\frac{\tau v_{\rm
W}}{2L},0,z_0)
+z_0\Phi(\frac{\tau v_{\rm W}}{2L},
\frac{L+2x}{2L},z_0)\big]\Big)
\end{eqnarray}
where $z_0\equiv\kappa_{-}/\kappa_{+}$, and
\begin{equation}
\Phi(x,y,z)\equiv\sum_{n=0}^{\infty}z^{2n}\ln\frac{x^2+(n+1+y)^2}{(n+1+y)^2}.
\label{eq:Phi}
\end{equation}
Evaluation of the Fourier integral $\int
d\tau(1-e^{i{\bar \omega}\tau})F(x,\tau)$
for arbitrary ${\bar \omega}$ is impossible in
this case. As ${\bar \omega}$ has now become a continuous variable
($\beta\to\infty$), however,  it is possible to find the asymptotics
of the integrand of Eq.~(\ref{eq:cond_corr})
for small ${\bar \omega}$ and then analytically continue to $i\omega$. Analysis
shows that in the limit ${\bar \omega}\to 0$  the main contribution to
 the Fourier integral  comes from the interval $\tau\geq
L/v_{\rm W}$, where the function $\Phi$ [Eq.~(\ref{eq:Phi})] can be
replaced by its large-$x$ asymptotic form: $\Phi(x,y,z)\approx \ln
x^2/(1-z^2)$.
It can also be shown that the imaginary
 part of $F_0-F_{{\bar \omega}}$ is asymptotically larger than its real part in
 the ${\bar \omega}\to 0$ limit. After simple manipulations one then obtains
 $F_{0}-F_{{\bar \omega}}=((L/L_{{\bar \omega}})^{\alpha}/i{\bar \omega})\times
I(\lambda)$, where
 $\lambda\equiv\omega_{\rm F}/{\bar \omega}$,
 \begin{equation}
 I(\lambda)\equiv\int_{0}^{\infty}dz\frac{z^{-\alpha}\sin
z}{\Big[1+\lambda^2z^2\Big]^{K_{\rm W}}}
 \label{eq:ilam}
 \end{equation}
 and $\alpha\equiv 2K_{\rm L}(K_{\rm L}-K_{\rm W})$. As we see, the exponent
$\alpha$, which is going
 to determine the $L$-scaling of $\overline{\delta g}$, depends both on
 $K_{\rm L}$ and $K_{\rm W}$ and is equal to the
 exponent of the $T$-scaling [i.e, to $2(1-K_{\rm W})$] only if $K_{\rm L}=1$.
Thus, the
 equivalence
 of $T$- and $L$-scalings exists only if the electrons in the leads
 do not interact. To be consistent with the physical content of the model,
 I will now consider only the case
 $K_{\rm L}=1$.
 We now need to evaluate the asymptotic
 behavior of the integral (\ref{eq:ilam}) for $\lambda\to\infty$. This
 can be done by using the Parseval formula for Mellin transforms
 \cite{bleistein}, according to which one has
 \begin{equation}
 I(\lambda)=\int_{c+i\infty}^{c+i\infty}\frac{ds}{2\pi i}
 \lambda^{-s}\Gamma(s/2)
 \Gamma({\bar s})\sin (\pi{\bar s}/2),
 \label{eq:parseval}
 \end{equation}
 where ${\bar s}\equiv 2K_{\rm W}-1-s$, and $c$ is chosen in the domain of
analyticity
 of the integrand ($0\leq {\rm Re} s\leq 2K_{\rm W}$).
 For $\lambda\to\infty$, the leading contribution to the sum over
 residues is given by the smallest second-order pole at $s=2K_{\rm W}$.
 I then find
 \begin{equation}
 I(\lambda)\Big|_{\lambda\to\infty}\approx\Gamma(K_{\rm W})\frac{\ln
 \lambda}{\lambda^{2K_{\rm W}}}.
 \label{eq:lamasy}\end{equation}
 In the limit being considered, $F_{0}-F_{{\bar \omega}}$ is $x$-independent,
and
 the integration over ${\bar x}$ in Eq.~(\ref{eq:cond_corr}) again gives $L$.
The continuation ${\bar \omega}\to i\omega-\epsilon$ can now be  performed.
Separating
the real and imaginary parts of $\ln\lambda$ in Eq.~(\ref{eq:lamasy}), I
finally  obtain the real part of the conductance in the form
\begin{equation}
\overline{{\rm Re}
g}=\frac{e^2}{h}-\frac{e^2}{h}2\pi^2\Gamma(K_{\rm
W})\frac{L}{\ell}\Big[\frac{2L\hbar\omega_{\rm F}}
{v_{\rm W}}\Big]^{2(1-K_{\rm W})}
\label{eq:res_low}
\end{equation}

Similar analysis can be performed in the case of a single potential barrier
inserted into the Luttinger-liquid wire which, in the case of a uniform
interaction strength,
was considered first in Refs.~\cite{fisher-prl,fisher-prb}. The result is
identical to the case of the extended random potential considered above:
in both the weak and strong disorder cases the
$T$- and $L$-dependences of the conductance are determined by the parameter
$K_{\rm W}$ of the wire.

The generalization for case of electrons with spin can readily be
performed. As in the homogeneous case \cite{fisher-prb,fukuyama},
the exponent $2-2K_{\rm W}$ is replaced by $2-K^{\rho}_{\rm W}-K^{\sigma}_{\rm
W}$, or, if the $SU(2)$ symmetry of the underlying Hubbard model is
preserved [i.e., $K^{\sigma}_{\rm
W}=1$] by $1-K^{\rho}_{\rm W}$, where $K^{\rho/\sigma}_{\rm W}$ are the
parameters of the charge/spin parts of the Luttinger-liquid in the wire.

This work was supported by the NSF under grant
DMR89-20538. I am grateful to P. Goldbart, E.
Fradkin, D. Loss, N. Sandler, and M. Stone for stimulating discussions.
P. Goldbart, N. Sandler, and M. Stone have also read the manuscript and
made valuable comments. P. Goldbart deserves special thanks
for attracting my attention to and providing me with a 	copy of
Ref.~\cite{bleistein}.


\begin{references}
\bibitem{fn_perturb} This assumption is adequate for the experimental
situation studied in Ref.~\cite{tarucha}, where $L$ was at least $\simeq 6$
smaller than the elastic mean free path of the unbounded 2DEG and the
total observed change in the conductance was $1-5\%$ of $e^2/h$ depending
on the wire length.
\end{references}
\end{document}